\documentclass[%
 aip,
 amsmath,amssymb,
 reprint,%
 citeautoscript, %
]{revtex4-1}

\usepackage{graphicx}
\usepackage{dcolumn}
\usepackage{bm}
\usepackage{float}

\begin{document}
\raggedbottom
\preprint{AIP/123-QED}

\title{Note: Vibronic coupling in light-harvesting complex II revisited}

\author{Eric A. Arsenault}
\altaffiliation{These authors contributed equally.}
\affiliation{\looseness=-1Department of Chemistry, University of California, Berkeley, CA 94720, USA}
\affiliation{\looseness=-1Kavli Energy Nanoscience Institute at Berkeley, Berkeley, CA 94720, USA}
\affiliation{\looseness=-1Molecular Biophysics and Integrated Bioimaging Division, Lawrence Berkeley National Laboratory, Berkeley, CA 94720, USA}
 
\author{Addison J. Schile}
\altaffiliation{These authors contributed equally.}
\affiliation{\looseness=-1Department of Chemistry, University of California, Berkeley, CA 94720, USA}
\affiliation{\looseness=-1Chemical Sciences Division, Lawrence Berkeley National Laboratory, Berkeley, CA 94720, USA}

\author{David T. Limmer}
  \email{dlimmer@berkeley.edu}
\affiliation{\looseness=-1Department of Chemistry, University of California, Berkeley, CA 94720, USA}
\affiliation{\looseness=-1Kavli Energy Nanoscience Institute at Berkeley, Berkeley, CA 94720, USA}
\affiliation{\looseness=-1Chemical Sciences Division, Lawrence Berkeley National Laboratory, Berkeley, CA 94720, USA}
\affiliation{\looseness=-1Materials Science Division, Lawrence Berkeley National Laboratory, Berkeley, CA 94720, USA}

\author{Graham R. Fleming}
  \email{grfleming@lbl.gov}
\affiliation{\looseness=-1Department of Chemistry, University of California, Berkeley, CA 94720, USA}
\affiliation{\looseness=-1Kavli Energy Nanoscience Institute at Berkeley, Berkeley, CA 94720, USA}
\affiliation{\looseness=-1Molecular Biophysics and Integrated Bioimaging Division, Lawrence Berkeley National Laboratory, Berkeley, CA 94720, USA}

\date{\today}

\maketitle

A growing body of work has pointed to vibronic mixing as a crucial design principle for efficient energy and charge transfer in natural\cite{fuller2014vibronic,dean2016vibronic,arsenault2020vibronic} and  artificial systems\cite{rozzi2013quantum,falke2014coherent}. Notable among these studies was the recent observation of vibronically-promoted ultrafast energy flow in the major antenna complex of green plants and algae light-harvesting complex II (LHCII)---the most abundant membrane protein on earth\cite{standfuss2005mechanisms}---via the emerging experimental technique two-dimensional electronic-vibrational (2DEV) spectroscopy.\cite{arsenault2020vibronic} This spectroscopy, which correlates electronic and nuclear degrees-of-freedom, shows promise for providing mechanistic insight into vibronic coupling, however, explicit theoretical input is necessary to extract such detail. In a separate paper\cite{arsenault2021vibronic}, we have developed a heterodimer model that describes various forms of vibronic coupling---which were speculated to be present in LHCII---resulting from diagonal electron-phonon coupling giving Franck-Condon (FC) activity and nuclear dependence of the electronic transition dipole moment giving Herzberg-Teller (HT) activity. Here, we draw connections between this theoretical work and recent experimental studies in order to demonstrate how HT activity is leveraged in the light-harvesting capabilities of LHCII.

\begin{figure}[H]
\begin{center}
\includegraphics[scale=0.495]{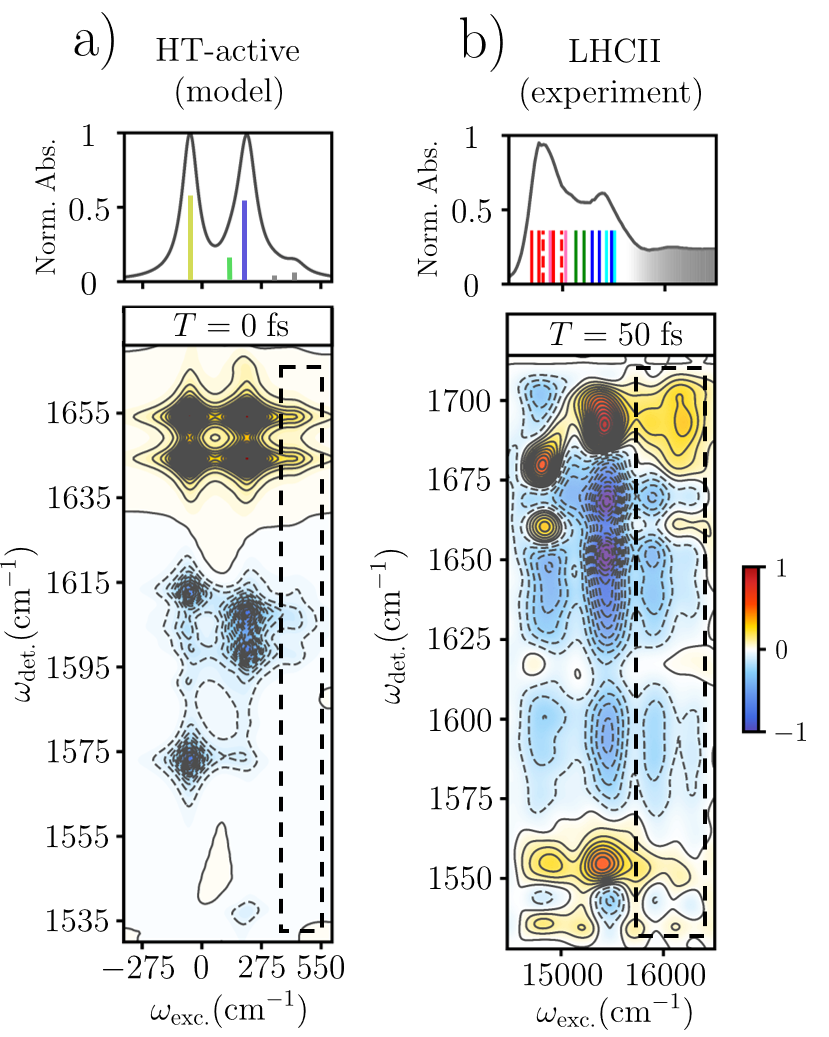}
\caption{\label{fig:note_fig1} Electronic linear absorption spectra (Top row) and 2DEV spectra (Bottom row) for a) HT-active multimode heterodimer model and b) LHCII at 77 K. Vertical lines in the linear absorption spectra denote excitonic states (see Ref. \citenum{novoderezhkin2005excitation} for LHCII). In the 2DEV spectra, the dashed box highlights the higher-excitation frequency portion of the spectra where HT-induced vibronic transitions appear. Positive features indicate ground state bleaches and negative features indicate excited state absorptions. b) has been adapted from Ref. \citenum{arsenault2020vibronic} with permission (http://creativecommons.org/licenses/by/4.0/).
}
\end{center}
\end{figure}

The striking resemblance between the static vibronic coupling signatures found in the simulated spectra and the experimental measurement---shown in Fig. \ref{fig:note_fig1}a and b, respectively---warrants the direct comparison we put forth here. While the model heterodimer mimics the linear absorption spectrum of LHCII, exhibiting two dominant peaks arising from the excitonic states of the electronically coupled chromophores, the presence of the higher-energy side-band is the most significant similarity. In the model, this feature emerged specifically as a result of HT activity\cite{arsenault2021vibronic}---remaining hidden when the system was only vibronically mixed by FC activity through a Huang-Rhys factor slightly larger than what would be expected for LHCII.\cite{jia1992simulations} This side-band also appeared in the heterodimer 2DEV spectra (highlighted by the dashed, black box in Fig \ref{fig:note_fig1}a) where the vibrational structure specifically replicated the excitonic state with significant electronic character from the higher-energy chromophore (analagous to chlorophyll $b$ in LHCII).\cite{arsenault2021vibronic} Remarkably, not only does the experimental 2DEV spectrum of LHCII clearly display side-band features, but the vibrational structure along these bands replicates that of the higher-lying excitonic states composed of mainly chlorophyll $b$ (spanning 15200$\sim$15600 cm$^{-1}$) as marked by the characteristic ground state bleach signature at 1690 cm$^{-1}$ and predicted by the model. These observations, in agreement with the key features displayed by the model, highlight the presence of HT activity in LHCII.

\begin{figure}[h]
\includegraphics[scale=0.5]{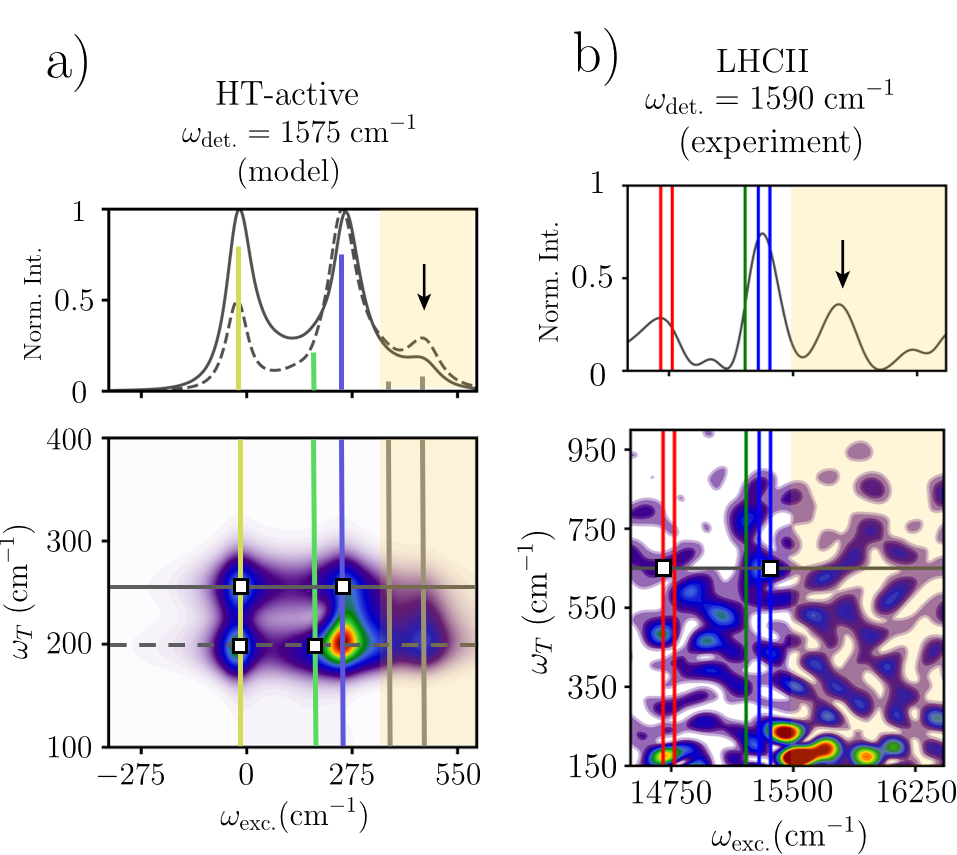}
\caption{\label{fig:note_fig2} Beat maps at a fixed detection frequency (Bottom row), $\omega_{\mathrm{det.}}$, for the a) model and b) experiment. Vertical lines denote excitonic states (for LHCII these are based on Ref. \citenum{novoderezhkin2005excitation}) and horizontal lines are specific beat frequencies slices which are shown explicitly above the beat maps. The yellow shaded region and black arrows indicate higher-energy vibronic states that are promoted by HT activity. At a given beat frequency, squares indicate the energetic gap between excitonic states corresponding to that frequency. 
b) has been adapted from Ref. \citenum{arsenault2020vibronic} with permission (http://creativecommons.org/licenses/by/4.0/)}
\end{figure}

2DEV spectra also evolve as a function of waiting time and, in this case, report on the energy transfer dynamics. In complex systems, where many quantum beating signals can be present during the waiting time, a frequency-domain characterization of these oscillatory signals is often performed and visualized with a beat map (see Ref. \citenum{arsenault2021vibronic} for details of this analysis), as shown in Fig. \ref{fig:note_fig2}. In 2DEV spectroscopy, these beat maps are typically calculated at a fixed, local detection mode which serves as a sensitive reporter of the interplay between excitonic states. The beat map from the model simulations (Fig. \ref{fig:note_fig2}a) displays two distinct dynamical frequencies that report on the two relevant energy gaps and importantly shows contributions from higher-lying excitonic states (shaded regions of Fig. \ref{fig:note_fig2}) at both dynamical frequencies. This manifests as an additional peak in the map that can be clearly seen at both beat frequencies (highlighted by black arrow in the top panel of Fig. \ref{fig:note_fig2}a). The participation of the higher-lying vibronic states at all beat frequencies was specific to HT activity.\cite{arsenault2021vibronic} While the experimental measurements (Fig. \ref{fig:note_fig2}b) are considerably more structured as a result of the many additional excitonic states relative to the model, the beat map shows a similar distribution of higher-energy vibronic activity across the numerous beat frequencies. One example of the prevalence of the high-energy vibronic states in the observed beat frequencies corresponding to excitonic energy gaps is presented in Fig. \ref{fig:note_fig2}b (see top panel).
Despite the complexity of the experimental beat maps, the general observable features from our model analysis elucidates the underlying mechanism of vibronic coupling in the energy transfer dynamics of LHCII, namely, the role of vibronic mixing in promoting additional, ultrafast relaxation pathways. 

More specifically, the model indicates that HT activity is the potential underlying mechanism of energy transfer. Further 2DEV studies on LHCII, featuring excitation well into the green absorption region\cite{arsenault2020role}, have indicated that vibronically-induced ultrafast energy flow persists across nearly 3000 cm$^{-1}$ of the electronic linear absorption spectrum of the complex, which in conjunction with the model results, suggests that HT activity is also the mechanism responsible for the extension and enhancement of the light-harvesting capabilities of LHCII across the photosynthetically active region.

E.A.A. and G.R.F. acknowledge support from the U.S. Department of Energy, Office of Science, Basic Energy Sciences, Chemical Sciences, Geosciences, and Biosciences Division. A.J.S. and D.T.L. were supported by the U.S. Department of Energy, Office of Science, Basic Energy Sciences, CPIMS Program Early Career Research Program under Award No. DE-FOA0002019. E.A.A. is grateful for support from the National Science Foundation Graduate Research Fellowship (Grant No. DGE 1752814). 

%

\end{document}